\documentclass[11pt,a4paper]{article}
\usepackage{authblk}
\usepackage{hyperref}
 
\usepackage{booktabs} 
\usepackage{url}
\usepackage{etoolbox}
\usepackage{enumitem}
\usepackage{graphicx}
\usepackage{multirow}

\newcommand{\N}{{\tt Note}}
\newcommand{\D}{{\tt Desc}}
\newcommand{\M}{{\tt Sum}}

\date{}

\begin{document}

\title{Benchmarking Clinical Decision Support Search}

\author{Vincent Nguyen, Sarvnaz Karimi, Sara Falamaki, Cecile Paris\\
CSIRO, Data61\\
Marsfield, NSW, Australia\\
\url{firstname.lastname@data61.csiro.au}}

\maketitle

\begin{abstract}
Finding relevant literature underpins the practice of evidence-based
medicine. From 2014 to 2016, TREC conducted a clinical decision support
track, wherein participants were tasked with finding articles relevant
to clinical questions posed by physicians. In total, 87 teams have
participated over the past three years, generating 395 runs. During
this period, each team has trialled a variety of methods. While there
was significant overlap in the methods employed by different teams,
the results were varied. Due to the diversity of the platforms used,
the results arising from the different techniques are not directly
comparable, reducing the ability to build on previous work. By using a
stable platform, we have been able to compare different document and
query processing techniques, allowing us to experiment with different
search parameters. We have used our system to reproduce leading
team's runs, and compare the results obtained.  By benchmarking our
indexing and search techniques, we can statistically test a variety of
hypotheses, paving the way for further research.\footnote{A system to run these experiments online will be made available to the research community.}
\end{abstract}

\section{Introduction}

Physicians are required to keep abreast of the latest medical
advances, as published in medical journals. At the time of writing,
over 27 million articles have been added to PubMed since Jan
2017~\cite{Pubmed}. To practice evidence based medicine, a clinician
must be able to locate relevant literature in a short time (less than
two minutes~\cite{Ely:Osheroff:1999}), clearly a difficult task. 
Since 2014, the Clinical Decision Support (CDS)
Track~\cite{Simpson:Vorhees:2014,Roberts:Simpson:2015,Roberts:Fushman:2016} 
at the Text Retrieval Conference (TREC) has provided a forum
for tackling this important issue. Each year, participants were
provided with a set of Electronic Health Records (EHRs) and
asked to find relevant articles. While verbose queries
have always been a challenge in information retrieval (IR), EHRs
pose a steeper challenge as these records are often riddled
with spelling, grammar, and punctuation errors as well as
medical jargon and abbreviations. While in 2014 and 2015 simulated
EHRs~\cite{Roberts:Simpson:2015} were used, in 2016 participants were
provided with real EHRs. These were arguably far more difficult to
process, resulting in a drop in recall and precision scores. In this
paper, we focus on the 2016 track (CDS'16).

While CDS tracks provide valuable datasets for evaluating search methodologies, it is difficult to compare different participants' algorithms. Since these platforms are seldom open sourced or made accessible to others, building on previous work is very challenging. While most participants use a combination of standard algorithms (e.g., seven out 24 teams used pseudo-relevance feedback), the results they obtain vary considerably~\cite{Roberts:Fushman:2016}.


To help further research in this area, we developed a platform
that allows us to compare methods used by the biomedical IR
community, in particular, TREC CDS participants. These methods
include query and document processing techniques, such as negation
detection, normalization, query expansion and reformulation, use of
knowledge databases, and learning to rank. Our goal is to facilitate
proof-of-concept approaches to answer the following questions using the
clinical search dataset provided by TREC: Given a clinical question,
what are the most promising retrieval methods? and; Does any specific
indexing method lead to better retrieval effectiveness?

Our experiments with a variety of methods identify some of the
most and least effective approaches.  We identify some of the
reproducible results from this track, and how our experimental results
compared. Our results form a benchmark for evaluating more sophisticated
algorithms. We identify some of the difficulties encountered in
attempting to reproduce competition results, and suggest how to mitigate
these problems in the future by accurately specifying the methods used.

\begin{table}[tb]
\begin{scriptsize}
\begin{center}
\caption{\label{tbl-summary}A brief summary of popular methods and search engines from different teams at TREC CDS 2016.}
\begin{tabular}{|l|p{0.65\linewidth}|}
\hline
\bf Method& \bf Number of teams (Team name)\\
\hline
Use of UMLS 	  & 8 (CBNU~\cite{cbnu:trec:2016}, CSIRO~\cite{Karimi:Falamaki:2016}, MerckKGaA~\cite{MERCKKGAA:trec:2016}, NCH-RISSI~\cite{nch_risi:trec:2016}, ECNU~\cite{ECNU:tre:2016}, DUTH~\cite{DUTH:trec:2016}, iRiS~\cite{iris:trec:2016}, SCAIAICLTeam~\cite{SCIAICLTeam:trec:2016})\\
\hline
Use of MeSH 	  & 8 (CBNU used level 2 MeSH headings related to diseases, ECNU, ETH~\cite{ETH:trec:2016}, MayoNLP~\cite{MayoNLPTeam:trec:2016}, IAII-PUT~\cite{IAII_PUT:trec:2016}, IRIT~\cite{IRIT:trec:2016}, NKU~\cite{NKU:trec:2016}, NLM-NIH~\cite{NLM_NIH:trec:2016})\\		   
\hline
Psuedo Relevance Feedback  & 7 (CBNU, DA-IICT~\cite{DA_IICT:trec:2016}, MerckKGaA, ETH, MayoNLP, NKU~\cite{NKU:trec:2016}, UNTIIA~\cite{UNTIIA:trec:2016})\\
\hline
Concept Extraction& 5 (CSIRO, DUTH, iRiS, IRIT, HAUT~\cite{HAUT:trec:2016})\\
(e.g., Metamap or MaxMatcher) & \\
\hline
Negation Detection& 3 (ETH - modified negated words found by NegEx to a new form, iRiS - removed negated terms found by NegEx, SCIAICLTeam - removed negated terms found by NegEx)\\
\hline
Word embeddings   & 3 (CBNU - source unknown, MerckKGaA - Wikipedia and \\
(e.g., Wikipedia or Medline)&  CDS 2016 corpus, ETH - CDS 2016 corpus)\\ 
\hline
Learning To Rank  & 3 (ETH, MerckKGaA, WHUIRGroup~\cite{WHUIRGroup:trec:2016})\\
\hline
Search engine & \\
\multicolumn{1}{|r|}{Terrier}& 6 (DA-IICT, ECNU, NKU, NLM-NIH, UNTIIA)\\
\multicolumn{1}{|r|}{Indri}  & 4 (MayoNLP, DUTH, iRis, WHUIRGroup)\\
\multicolumn{1}{|r|}{Solr}	 & 3 (CBNU, CSIRO, MerckKGaA)\\
\multicolumn{1}{|r|}{Elastic Search}& 2 (NCH-RISSI, HAUT)\\
\multicolumn{1}{|r|}{Lucene} & 2 (CCNU, SCAIAICLTeam)\\
\hline
\end{tabular}
\end{center}
\end{scriptsize}
\end{table}

\section{Related Work}

\subsubsection*{Medical Information Retrieval} TREC has a
long history of running tracks in the medical domain, including the
Genomics track (2003-2007), the Medical track on electronic health
records (2011-2012), and the TREC Clinical Decision Support (CDS) track
which ran for three years (2014-2016). The TREC CDS track aimed
to facilitate the answering of clinical questions pertaining to better
patient care~\cite{Roberts:Simpson:2016}. 
After the results are released, most teams publish a short report
detailing their search methodology and results. 

A high-level list of approaches taken by 2016 CDS participants is
summarized in Table~\ref{tbl-summary}. While some teams implemented
unique methods, there was still a large overlap in methods used. For
example many teams identified UMLS concepts in articles and topics
using Metamap. We have implemented some of these popular methods and
benchmarked them on our platform. Since not all the teams detailed
the exact nature of the search algorithm or indexing engine used
in their experiments, Table~\ref{tbl-summary} only summarizes
those who clearly documented their method. The top performing
team of 2016 (FDUDMIIP) did not report on their methodology. We
attempted to reproduce MerckKGaA~\cite{Guru:Bauer:2016} results,
as they ranked second in 3 out of 4 metrics. Our methodology is
explained in Section~\ref{sec-methods}.  We also report on individual
methods from other teams who utilized MeSH, negation detection using
NegEX by removing negated terms (iRiS~\cite{iris:trec:2016} and
SCIAICLTeam~\cite{SCIAICLTeam:trec:2016}) and concept extraction using
Metamap. We note that similar methods were also examined in the 2015 CDS
track~\cite{Roberts:Simpson:2015}.

\subsubsection*{Evaluation in Information Retrieval} Lack of comparable
results in information retrieval have been observed and investigated in
the IR community. A brief list of evaluation issues using TREC like test
collections can be found here~\cite{Scholer:Kelly:Carterette:2016}. We
are not focusing the test collections creation issue in this
work. While having widely accessible test collections is paramount
in evaluating one's IR solution, we cannot compare results obtained
by different research groups without a unified platform and
baseline benchmarks. We note that in 2009 Armstrong {\em et.
al.}~\cite{Armstrong:2009} investigated the problem of reporting
improvements made over weak baselines in the ad hoc retrieval process
tested in a TREC setting~\cite{Armstrong:Moffat:2009}. Unfortunately
{\tt EvaluatIR}, the platform Armstrong proposed for comparing different
IR systems\footnote{{\tt EvaluatIR} at \url{www.evaluatIR.org}} is no
longer publicly accessible. EvaluatIR allowed researchers to upload
the output of their systems and have them evaluated and compared
against baselines. Another, more recent platform, is provided by
EvALL~\cite{EvALL:2017}. In this system some of the existing shared
tasks are benchmarked, and new benchmark data can be uploaded.  Inspired
by these systems, we created a platform that allows the testing of a
variety of retrieval methods on the CDS'16 corpus. While EvALL is a
generalised platform, we focus solely on biomedical IR, with its unique
challenges and methods, which include dealing with medical ontologies
such the UMLS.
\section{Dataset and Indexing} 
The corpus provided by CDS'2016 is a
snapshot of all published medical literature from PubMed Central taken
on 28 March 2016. It contains 1.25 million full-text journal articles,
excluding their references, keywords and MeSH headings. These documents
are encoded in NXML format (an XML format extended using
National Library of Medicine (NLM) journal archiving and interchange tag
library). After re-encoding each document into ASCII text, we indexed it
using Solr~\cite{Solr}, the same search engine that MerckKGaA, CBNU,
and CSIRO teams used. At the time of indexing, we appended MeSH keywords
(as published in the corresponding Medline abstracts) and generated 
Metamap concepts for each article.

Each year, TREC CDS has provided 30 topics to generate queries with. The
2016 task provided topics based on nursing admission notes. Each topic
had three different fields: (1) Note (\N) or the original clinical
note; (2) Description (\D), a simplified version of note where all
abbreviations and jargon were removed; and, (3) Summary (\M), a
condensed version of the description removing all the irrelevant
information.

\section{Methods\label{sec-methods}}

We investigate three sets of approaches common amongst the CDS'16
submissions: (1) use of knowledge bases; (2) query expansion and
reformulation; and (3) application of natural language processing
techniques, such as negation detection. Some of these techniques (chosen
from the systems in Table~\ref{tbl-summary}) are explained below.

\subsubsection*{Normalizing Demographics} 
Clinical notes used as topics
contain references to patient demographics, including age, sex and
cohort information. We apply a set of regular expressions to the topics
in order to normalize age and gender references as detailed by the CSIRO
team~\cite{Karimi:Falamaki:2016}. For example {\em 86 y/o m} is replaced
with {\em elderly male}.

\subsubsection*{Handling Negation}
There are two main methods of dealing with negation in text: (1) removing negated terms; and (2) changing them to a
unified term, for example ``no pain'' becomes ``no-pain''. We chose the first approach. Using Metamap Lite's~\cite{DemnerFushman:Rogers:2017}
NegEx algorithm, negated terms were identified and removed from
the topics. This method was used by iRiS~\cite{iris:trec:2016} and SCIAICLTeam~\cite{SCIAICLTeam:trec:2016}. However, not all
the details in SCIAICLTeam's system were clear. That is, while they use
Metamap's negation detection module, they also apply a rule-based text
pre-processing step which transforms text based on a dictionary they
have created (not cited, not shared). For example, they replace 
the term ``status post'' with ``after''.

\subsubsection*{Concept Extraction}
We use Metamap to extract medical concepts from both the topics and
the documents and assign them to UMLS concepts. Since there are a large
number of semantic types defined in the UMLS metathesaurus, many
of them irrelevant to our task, we re-implement what MerckKGaA
team~\cite{Guru:Bauer:2016} reported and only extract concepts with
the following semantic types: Disease or Syndrome, Sign or Symptom,
Pathologic Function, Diagnostic Procedure, Anatomical Abnormality,
Laboratory Procedure, Pharmacologic Substance, Neoplastic Process, and
Therapeutic or Preventive Procedure. 

\subsubsection*{Faceted Search} 
Scientific articles often follow a defined structure by including
different {\em fields} or {\em entities} such as title and author. We
use these fields as {\em facets}, in two ways: (1) filtering the index,
and (2) weighting facets. We filter the index by using the index of one
facet at a time. This is done to assign weights based on the importance
of certain parts of a document. We experimented with a range of $[0,2]$
for weights. The results we report here were the optimal weights found
empirically.

To find the optimal weights for each facet we used a hill
climbing algorithm, inspired by the work by the WSU-IR
team~\cite{Balaneshin:Kordan:2016,kordan:kotov:2015}. We weight
different facets by assigning a relative weight determined by
normalizing the infNDCG score. The hill climbing algorithm stores the
current global minimum, and if it is caught in a local minimum, it
will probe the feature space for a more suitable minimum using random
restart. Graduated optimization was not included in the algorithm
because the feature hyperspace was considered discrete. This algorithm
was run for approximately 30 epoch for each query type with the current
global minimum being passed at every epoch. No team at 2016 track
implemented this method. However, given the success of WSU-IR in the
CDS Track in 2015, we included this experiment to evaluate its 
performance on our platform.

\subsubsection*{Query Expansion using Pseudo-Relevance Feedback and Word Embeddings}
Two query expansion techniques-- Pseudo-Relevance Feedback (PRF), and
query expansion using semantically similar words extracted using Word
Embeddings (WE)---were investigated by ten different teams (refer to
Table~\ref{tbl-summary}). For PRF, we experimented adding N words
using the top-10 to top-40 retrieved documents. We also experimented
with weights assigned to the expanded terms (zero to one). The optimal
weights for PRF were 0.8 for \D, 0.2 for \M, and 0.9 for \N. To find
the upper bound of what we can achieve, we also implemented Relevance
Feedback (RF); that is, only relevant documents retrieved in the top-30
results were used to expand the queries (Top-30 was suggested by
MerckKGaA).

Word embeddings were created using a combination of Wikipedia and
Medline abstracts using Gensim. We only added a maximum of three
semantically related words from word embeddings for each word of
the queries with an upper limit of 40 words for the entire query as
suggested by~\cite{Bendersky:Croft:2009}.

\subsubsection*{Learning to Rank}
A popular approach for improving retrieval effectiveness is Learning-to-Rank (LTR). In TREC CDS, some of the high scoring teams, such as ECNU team~\cite{Song:He:Hu:2015} and MerckKGaA~\cite{Guru:Bauer:2016}, applied LTR techniques for re-ranking. MerckKGaA used LambdaRank~\cite{Guru:Bauer:2016} with the following features: BM25 scores from PRF (with and without UMLS query expansion), document distances between topic and document titles using word embeddings, articles types and type of the topic as treatment, diagnosis and test. We have also implemented this method. We used topics and relevance judgments from the 2014 and 2015 tracks as training data. Our features were scores from BM25, PRF, and topic types, document distances between topic and document titles using word embeddings as well as topic category (\N, \D, or \M).
\begin{table}[tb]
\begin{center}
\begin{scriptsize}
\caption{\label{tbl-methods}Comparison of different ranking strategies.}
\tabcolsep 1pt
\begin{tabular}{llc lc lc lc lc lc lc lc}
\toprule
\multicolumn{2}{c}{\bf Method}&&\bf Query&& \bf infNDCG && \bf infAP && \bf R-Prec && \bf P@10\\
\midrule
Baseline & BM25 && \N &&  0.1074 &&  0.0079 &&  0.0749 &&  0.1400 \\  
		 & && \D && 0.1067 &&  0.0060 &&  0.0766 &&  0.1200 \\
		 & && \M &&  0.1721 &&  0.0158 &&  0.1167 &&  0.2067 \\
\midrule
Negation &   	&& \N && 0.1104 && 0.0088 && 0.0772 && 0.1467\\ 
Detection& 		&& \D && 0.1097 && 0.0062 && 0.0766 && 0.1233\\
		 & 		&& \M && 0.1726 && 0.0159 && 0.1171 && 0.2067\\
\midrule
Demographics& &&\N && 0.1102 && 0.0089 && 0.0778 && 0.1400 && \\  
		& && \D && 0.1107 && 0.0064  &&0.0772 && 0.1267\\
		& && \M && 0.1755 && 0.0166 && 0.1149 && 0.2200\\
\midrule
Filtering&Title&& \N && 0.0727 && 0.0105 && 0.0393 && 0.0833\\
Facets& && \D && 0.0785 && 0.0055 && 0.0389 && 0.0867\\
			 & && \M && 0.1262 && 0.0134 && 0.0669 && 0.1367\\
		\cline{2-14}
		&Title&& \N && 0.0712 && 0.0099 && 0.0374$^\ddagger$ && 0.0833\\
		&+Concepts && \D && 0.0808 && 0.0056 && 0.0362 $^\ddagger$ && 0.0767\\
		& && \M && 0.1039 && 0.0086 && 0.0561$^\dagger$  && 0.1100$^\dagger$ \\
		\cline{2-14}
		&Abstract&& \N && 0.0788$^\ddagger$  && 0.0050 $^\dagger$ && 0.0475 $^\ddagger$ && 0.1233\\
		& && \D && 0.0841$^\dagger$  && 0.0047 && 0.0477$^\ddagger$  && 0.0867\\
		& && \M && 0.1201 && 0.0108 && 0.0775$^\ddagger$  && 0.1500$^\ddagger$ \\
		\cline{2-14}		
		&Abstract  &&\N && 0.0769$^\ddagger$  && 0.0051$^\dagger$  && 0.0469 $^\ddagger$ && 0.1067\\
		&+Concepts && \D && 0.0827$^\dagger$  && 0.0046 && 0.0467 $^\ddagger$ && 0.0967\\
		&	   && \M && 0.1268$^\dagger$  && 0.0128 && 0.0820$^\ddagger$  && 0.1700\\
		\cline{2-14}		
		&Title&& \N && 0.0800$^\ddagger$  && 0.0052 $^\ddagger$ && 0.0491 $^\ddagger$ && 0.1467\\
		&+Abstract  && \D && 0.0913 && 0.0050 && 0.0511 $^\ddagger$  && 0.1000\\
		& && \M && 0.1316$^\dagger$  && 0.0126 && 0.0786$^\ddagger$  && 0.1533 $^\dagger$ \\
		\cline{2-14}		
		&Title+Abstract&&\N && 0.0783 $^\ddagger$  && 0.0053 $^\dagger$ && 0.0497 $^\ddagger$ && 0.1133\\
		&+Concepts && \D && 0.0823 $^\dagger$  && 0.0046 && 0.0496 $^\ddagger$ && 0.0967\\
		&	   && \M && 0.1356 $^\dagger$  && 0.0129 && 0.0796 $^\ddagger$  && 0.1833\\
		\cline{2-14}		
		&Body   && \N && 0.1079 && 0.0073 $^\ddagger$ && 0.0730 && 0.1367\\
		& && \D && 0.1024 && 0.0055 && 0.0691 $^\ddagger$ && 0.1200\\
		& && \M && 0.1541 $^\ddagger$  && 0.0131 $^\dagger$ && 0.1045 $^\ddagger$ && 0.1933\\
\midrule
Weighting&Title&& \N && 0.1228 && 0.0119 && 0.0821 && 0.1833$^\dagger$\\ 
Facets		 & && \D && 0.1286 && 0.0084 && 0.0806 && 0.1567\\
			 & && \M && 0.1738 && 0.0190 && 0.1106  && 0.2433\\
		\cline{2-14}		
		&Title	   && \N && 0.1241 && 0.0150 && 0.0832 && 0.1733\\
		&+Concepts && \D && 0.1284 && 0.0090 && 0.0779 && 0.1667\\
		&	  && \M && 0.1620 && 0.0197 && 0.1075 && 0.2367\\
		\cline{2-14}
		&Abstract && \N && 0.1061 && 0.0084 && 0.0718 && 0.1700\\
				& && \D && 0.1027 && 0.0060 && 0.0730 && 0.1333 \\
				& && \M && 0.1540 && 0.0145 && 0.1047 && 0.2000\\	
		\cline{2-14}		
		&Abstract  && \N && 0.1031 && 0.0091 && 0.0707 && 0.1633\\
		&+Concepts && \D && 0.0990 && 0.0056 && 0.0688 && 0.1400\\
		& 	   && \M && 0.1505 && 0.0152  && 0.1051 && 0.2200 \\
		\cline{2-14}	    
		&Title	   && \N && 0.1111 && 0.0086 && 0.0738 && 0.1767\\
		&+Abstract && \D && 0.1057 && 0.0059 && 0.0740 && 0.1233\\
		& 	       && \M && 0.1603 && 0.0152 && 0.1021 && 0.2000\\
		\cline{2-14}
		&Title+Abstract && \N && 0.1045 && 0.0085 && 0.0690 && 0.1733\\
		&+Concepts && \D && 0.1088 && 0.0082 && 0.0732 && 0.1367\\
		& 		   && \M && 0.1567 && 0.0176 && 0.1029 && 0.1900\\		
		\cline{2-14}	    
		&Body     && \N && 0.1038 && 0.0073 && 0.0738 && 0.1267\\
				& && \D && 0.1011 && 0.0052$^\ddagger$ && 0.0692$^\ddagger$ && 0.1233\\
				& && \M && 0.1547 && 0.0135$^\ddagger$ && 0.1078$^\ddagger$ && 0.1967\\
		\cline{2-14}  
     &Hill	   &&\N && 0.1601$^\ddagger$  && 0.0148$^\dagger$  && 0.0904 $^\dagger$ && 0.2300$^\dagger$  \\
	 &climbing &&\D && 0.1738$^\ddagger$  && 0.0177$^\ddagger$ && 0.0913 && 0.2000$^\ddagger$ \\ 
   	 &		   &&\M && 0.2704$^\ddagger$  && 0.0395$^\ddagger$ && 0.1556$^\ddagger$ && 0.3133$^\ddagger$  \\  
	 \cline{2-14} 
     &Hill	   &&\N && 0.1666$^\ddagger$ && 0.0149 $^\dagger$ && 0.0911 &&0.2300$^\ddagger$ \\
	 &climbing &&\D && 0.1746$^\ddagger$ && 0.0178$^\ddagger$ && 0.0907 &&0.2033$^\ddagger$\\
   	 &	w/ negation   	   &&\M && 0.2777$^\ddagger$ && 0.0398$^\ddagger$ && 0.1569$^\ddagger$ &&0.3167$^\ddagger$\\   
		\cline{2-14}  
	&MeSH&&\N && 0.1668$^\ddagger$ && 0.0149$^\dagger$ && 0.0911 && 0.2300$^\ddagger$ \\
	& Filtered &&\D && 0.1718$^\ddagger$ && 0.0178$^\ddagger$ && 0.0907 && 0.2033$^\ddagger$ \\
	&     		  &&\M && 0.2111$^\ddagger$ && 0.0277$^\ddagger$ && 0.1429$^\ddagger$ && 0.2600$^\dagger$ \\  
\bottomrule
\end{tabular}
\end{scriptsize}
\end{center}
\end{table}

\section{Experiments} \subsubsection*{Experimental Setup} In
all our experiments, we used the Porter Stemmer and removed
stopwords. For evaluations, we used the four metrics proposed
in the CDS track: infNDCG (inferred NDCG)~\cite{Yilmaz:2008},
infAP (inferred average precision)~\cite{Yilmaz:2006}, R-prec
(Recall Precision) and P@10 (Precision at rank 10), with infNDCG
being the main metric. The significance of improvements over
the baseline is tested using a paired 2 sample $t$-test and
is represented in two scales of 95\% and 98\% confidence.
\subsubsection*{Results} Table~\ref{tbl-methods} compares a basic {\em
baseline}---BM25 with no query expansion or other topic or document
preprocessing---with other methods. The baseline uses the Solr {\em
eDisMax} function for query processing; no weights are used.  While
performing negation detection on its own did not improve the results
significantly, using it in conjunction with other techniques in the hill
climbing, word embeddings and PRF+UMLS runs (Table~\ref{tbl-QEmethods})
produced statistically significant improvements. Normalizing
demographics did not lead to a statistically significant improvement.

We used the second set of experiments, {\em filtering facets},
to estimate how much each facet contributes to finding relevant
results.  Predictably, searching individual facets resulted in a
drop in all four metrics for all facets. We found that the body of
the articles contributes to retrieving over 50\% of the relevant
results. Additionally: (1) adding UMLS concepts did not improve
retrieval using titles only; and (2) concepts in abstracts slightly
improved retrieval for \D$\;$ and \M, but not \N. Initially, we gave
each individual facet a weight of 2. We observed a statistically
significant improvement for P@10 for title (+3.9\%). The infAP and
R-Prec metrics were degraded for the article body. All other metrics
showed a statistically insignificant improvement. We then experimented
with hill climbing to obtain optimal weights for all facets. Using these
weights increased scores across the board statistically significantly.

Table~\ref{tbl-QEmethods} compares popular query expansion methods,
including PRF, WE, as well as using lexicons such as MeSH and UMLS. Our
best runs for PRF are expanded with words from the top-30 documents. The
results showed significant improvements over the baseline for all three
types of the queries. The last set of results belong to LTR. For \N$\;$
and \M$\;$ we observed significant improvements over PRF, PRF+UMLS+WE,
and the baseline.


\subsubsection*{Comparison to Other Systems} We compare our results with CBNU~\cite{cbnu:trec:2016} and
CSIRO~\cite{Karimi:Falamaki:2016} baseline runs and our replication of the
MerckKGaA~\cite{MERCKKGAA:trec:2016} runs. CBNU reported their baseline
to be Solr BM25. Their reported results for infNCG, R-Prec and P@10 for
both \N$\;$ and \M$\;$ are higher than our BM25 run. For example, they report
0.1927 infNDCG for \M$\;$ while ours is 0.1721. They do not
disclose their Solr version or the preprocessing they
do on the queries and documents. Interestingly, their \N$\;$ run
results for infNDCG and P@10 match exactly with our Weighting Facets
run with title boosted (Table~\ref{sec-methods}, first row, forth
section). This suggests that there must be some unreported parameters set in
their system. Our baseline runs are all above
the CSIRO reported runs, due their index being incomplete at submission time.

MerckKGaA reported four runs: \N$\;$ with PRF, \M$\;$ with PRF, \M$\;$ with
PRF+UMLS, and \M$\;$ with PRF+UMLS+LTR. They did not disclose their
preprocessing steps, and used an older version of Solr (5.5.2 vs
6.4.1). 

Using the same weights as they report for a PRF only run, we achieve
an infNDCG of 0.1234, lower than their score of 0.1504. Increasing the
weights to 0.2 improve our infNDCG results (0.1515). We achieve similar
results in the \M$\;$ PRF run: MerckKGaA 0.2223, we scored 0.2111 with 0.1
weight and 0.2161 with a weight of 0.2.

The PRF+UMLS run by MerckKGaA was constructed using PRF using by
extracting UMLS concepts from the initial query, and appending them to
it.  By doing this, they improved infNDCG of their PRF-only run from
0.2223 to 0.2261.

In our PRF+UMLS runs, we report infNDCG of 0.2042. We used PRF on the initial
query and then expanded it with UMLS concepts. We experimented with
another run (PRF+UMLS+WE) where we also expand the queries using
terms suggested by word embeddings. This method improved PRF 
infNDCG slightly, from 0.2161 to 0.2215. This is 
similar to the upward trend reported by MerckKGaA. Table~\ref{tbl-QEmethods} 
lists MerckKGaA's best run (\M$\;$ with
PRF+UMLS+LTR). Our LTR run has a lower infNDCG (by 0.0229) and exactly
the same P@10. The differences are related to the features used
in re-ranking. Although they listed all the features they
used, it was not clear how the product of the features (BM25 and WE) were calculated. As a result, we dropped that feature set.

\begin{table}[tb]	
\begin{center}
\begin{scriptsize}
\caption{\label{tbl-QEmethods}A comparison of query expansion and re-ranking techniques. RF: relevance feedback and WE: expansion using word embeddings.}
\tabcolsep 2pt
\begin{tabular}{lc lc lc lc lc lc lc lc}
\toprule
\bf Method&&\bf Query&& \bf infNDCG && \bf infAP && \bf R-Prec && \bf P@10\\
\midrule
PRF&&\N && 0.1516$^\ddagger$ && 0.0133$^\dagger$ && 0.0864 && 0.1700\\
&&\D && 0.1520$^\ddagger$ && 0.0128$^\ddagger$ && 0.0910 && 0.1833$^\ddagger$\\
   &&\M && 0.2169$^\ddagger$ && 0.0261$^\ddagger$ && 0.1439$^\ddagger$ && 0.2733$^\dagger$\\
\midrule
PRF&&\N && 0.1515$^\ddagger$ && 0.0131$^\dagger$ && 0.0866  && 0.1700\\
w/ negation   &&\D && 0.1559$^\ddagger$ && 0.0130$^\ddagger$  && 0.0919 && 0.1833$^\ddagger$ \\
   &&\M && 0.2161$^\ddagger$ && 0.0260$^\ddagger$ && 0.1437$^\ddagger$  && 0.2700 $^\dagger$ \\
\midrule
WE 		&&\N && 0.1006 && 0.0070 && 0.0775 && 0.1233 \\
       &&\D && 0.1147 && 0.0070 && 0.0758 && 0.1200 \\
      &&\M && 0.1777 && 0.0172 && 0.1227$^\ddagger$ && 0.2233 \\
\midrule   
WE &&\N && 0.1020$^\dagger$ && 0.0131 && 0.0866 && 0.1700 \\
w/ negation   &&\D && 0.1203 && 0.0082 && 0.0772 && 0.1500 \\
   &&\M && 0.1788 && 0.0174 && 0.1246$^\ddagger$ && 0.2267 \\
\midrule
UMLS&&\N && 0.1202$^\dagger$ && 0.0093 && 0.0814 && 0.1733$^\ddagger$\\
 &&\D && 0.1203 && 0.0086 && 0.0845 && 0.1167\\
    &&\M && 0.1757 && 0.0176 && 0.1202 && 0.2267\\
\midrule
UMLS&&\N && 0.1190$^\dagger$ && 0.0089 && 0.0822$^\dagger$ && 0.1733$^\ddagger$\\
w/ negation    &&\D && 0.1211 && 0.0087 && 0.0862 && 0.1300\\
    &&\M && 0.1750 && 0.0175 && 0.1203 && 0.2233\\
\midrule
PRF+UMLS   &&\N && 0.1496$^\ddagger$ && 0.0134$^\ddagger$ && 0.0915$^\dagger$ && 0.1700\\
 		   &&\D && 0.1519$^\ddagger$ && 0.0131$^\ddagger$ && 0.0954$^\dagger$ && 0.1900$^\ddagger$\\
  		   &&\M && 0.2042$^\dagger$ && 0.0273$^\ddagger$ && 0.1382$^\dagger$ && 0.2567\\
\midrule 
PRF+UMLS    &&\N && 0.1500$^\ddagger$ && 0.0133$^\ddagger$ && 0.0909$^\dagger$ && 0.1733\\
w/ negation &&\D && 0.1541$^\ddagger$ && 0.0134$^\ddagger$ && 0.0953$^\dagger$ && 0.1900$^\ddagger$\\
  		    &&\M && 0.2052$^\dagger$ && 0.0273$^\ddagger$ && 0.1383$^\dagger$ && 0.2567\\
\midrule 
PRF+UMLS+WE&&\N && 0.1421$^\ddagger$ && 0.0126$^\ddagger$ && 0.0920$^\ddagger$ && 0.1833\\
 		   &&\D && 0.1504$^\ddagger$ && 0.0128$^\ddagger$ && 0.0952$^\dagger$ && 0.1800$^\ddagger$\\
  		   &&\M && 0.2215$^\ddagger$ && 0.0274$^\ddagger$ && 0.1470$^\ddagger$ && 0.2633$^\dagger$\\
\midrule 
PRF+UMLS+WE&&\N && 0.1426$^\ddagger$ && 0.0127$^\ddagger$ && 0.0922$^\ddagger$ && 0.1767\\
w/ negation&&\D && 0.1517$^\ddagger$ && 0.0128$^\ddagger$ && 0.0950$^\dagger$ && 0.1800$^\ddagger$\\
  		   &&\M && 0.2187$^\ddagger$ && 0.0269$^\ddagger$ && 0.1484$^\ddagger$ &&0.2600$^\dagger$\\
\midrule    
RF  &&\N && 0.1692$^\ddagger$ && 0.0191$^\ddagger$  && 0.1003$^\ddagger$&& 0.2567\\
	&& \D && 0.1692$^\ddagger$ && 0.0182$^\ddagger$ && 0.1105$^\ddagger$ && 0.2400$^\ddagger$\\ 
    &&\M && 0.2324$^\ddagger$ && 0.0337$^\ddagger$ && 0.1557$^\ddagger$&& 0.3700$^\dagger$\\     
\midrule     		  
LTR	&& \N && 0.1667$^\ddagger$ && 0.0135$^\ddagger$ && 0.1061$^\dagger$ && 0.2229$^\ddagger$\\		
	&& \D && 0.1750$^\ddagger$ && 0.0924$^\ddagger$ && 0.0995$^\dagger$ && 0.1989$^\ddagger$\\
	&& \M && 0.2264$^\ddagger$ && 0.0301$^\ddagger$ && 0.1590$^\ddagger$&& 0.3530$^\ddagger$\\  
\midrule
MerckKGaA 	&& \M && 0.2493 && 0.0315&& 0.1744 && 0.3500\\ 
\midrule     		  
TREC Median	&& \N && 0.1228$^\ddagger$ && 0.0099$^\ddagger$ && 0.0792 && 0.1833 \\ 	
	&& \D && 0.1043 && 0.0065 && 0.0648 && 0.1533 \\
	&& \M && 0.1859 && 0.0196 && 0.1220 && 0.2633 \\  
\bottomrule
\end{tabular}
\end{scriptsize}
\end{center}
\end{table}

\subsubsection*{Query Analysis} 
\begin{figure}[tb]
\begin{center}
\tabcolsep 0pt
\begin{tabular}{cc}
\N & \M\\
\includegraphics[scale=0.09]{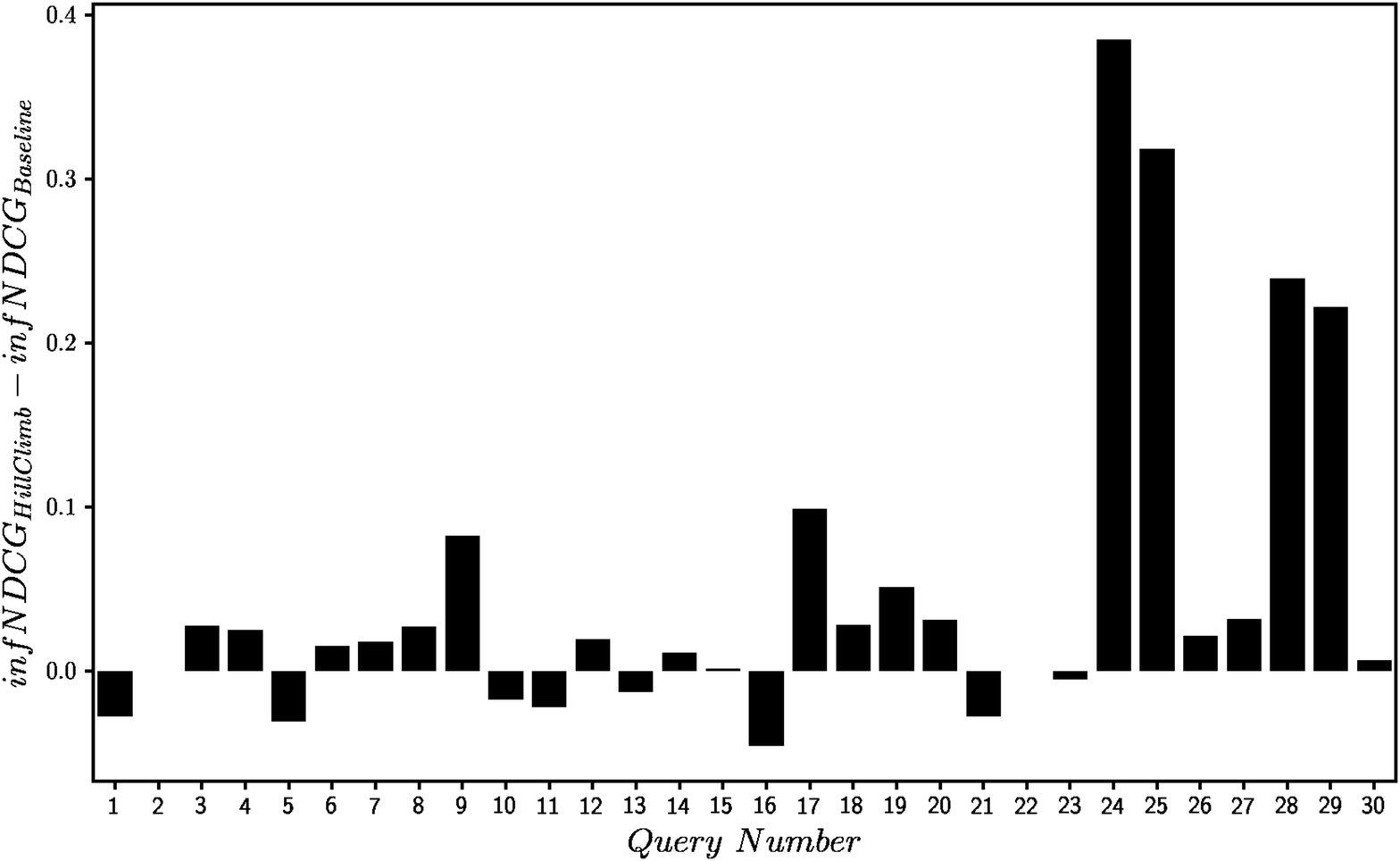}&\includegraphics[scale=0.09]{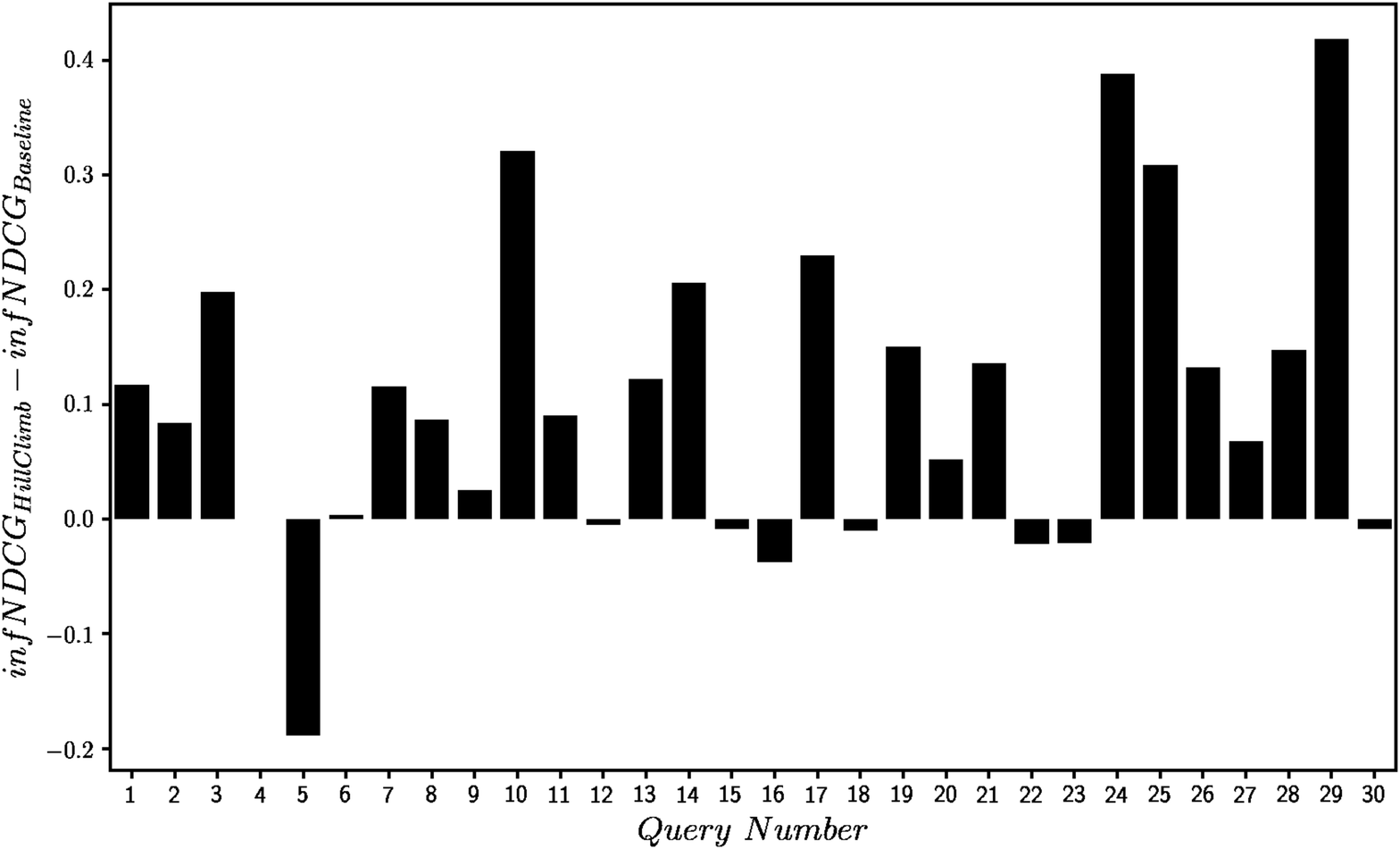}\\
\D &\\
\includegraphics[scale=0.09]{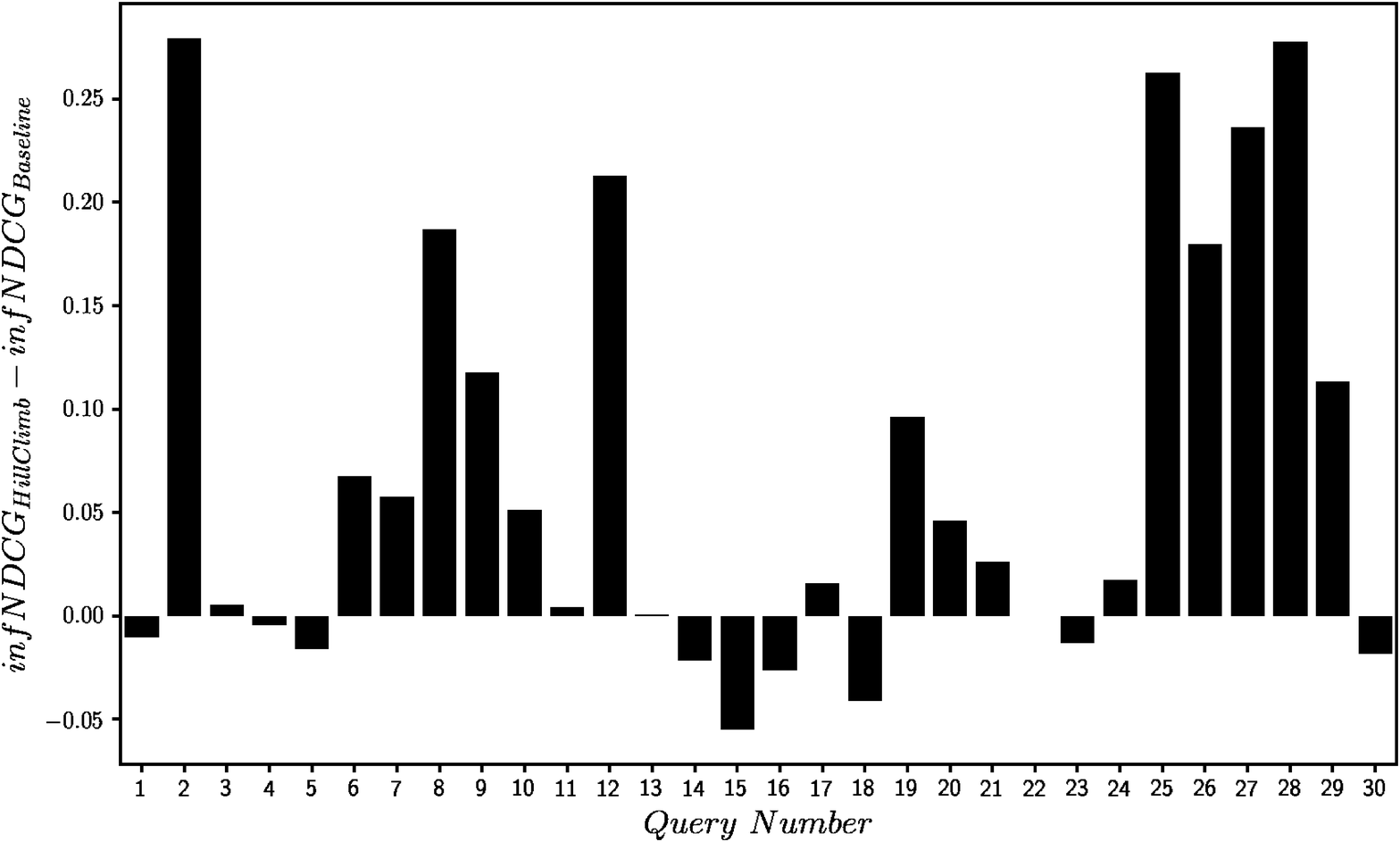}&\\
\end{tabular}
\caption{\label{fig-perquery}Per query changes to infNDCG using the hill climbing approach.}
\end{center}
\end{figure}

To identify where our best method, hill
climbing, improved queries, we show the differences in infNDCG between the
baseline and hill climbing run in Figure~\ref{fig-perquery}. Changes
for \D$\;$ were similar to \N$\;$ and therefore not shown. We observed very
different changes with \N and \M. For \M$\;$ we observed improvements
for 20 out of 30 queries. For \N, only 17 queries were positively
impacted. This emphasizes the difficulty of working with notes as
topics.

\begin{figure}[tb]
\begin{center}
\includegraphics[scale=0.2,angle = -90]{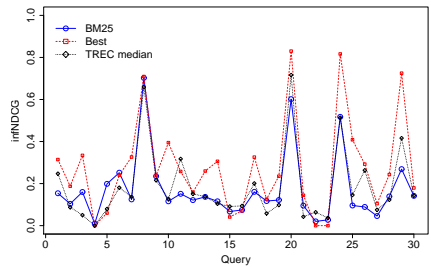}
\caption{\label{fig-query}Query level comparison of BM25, our best PRF, and TREC median for \M.}
\end{center}
\end{figure}

We compared the results of our PRF run and baseline BM25
with the official TREC 2016 CDS median for 30 queries in
Figure~\ref{fig-query}. The hardest topic was 22 (summary: {\em 94 M with CAD
s/p 4v-CABG, CHF, CRI presented with vfib arrest./ type: treatment)},
since there were only 8 relevant documents. The best run using hill
climbing did not retrieve any of the relevant documents (not even in
the top 1000). The baseline had three of these documents in the top
1000, one in the top 100, and none in the top 10. This particular topic
was high on vital statistics, but low on other information, which made
finding relevant information difficult.

\section{Conclusions} 

TREC Clinical Decision Support (CDS) track has been running since 2014 providing an opportunity for the IR community to investigate
ways to search biomedical literature to improve patient care. This track is popular, with many teams participating. However, most of the
runs are not conclusive as to whether or not the proposed approach is effective. 

In order to perform a comprehensive assessment of different search
methodologies, we developed a platform that allows the user to formally
specify the methods they use, re-run past experiments, and analyze the
findings based on qrels provided by the TREC CDS team. We have benchmarked the most common methods on their own
as well as combinations of these methods. The methods include: query
and document expansion using UMLS concepts, word embeddings, negation
detection and removal, and LTR.

While the exact results could not be reproduced due to lack of
sufficient details on preprocessing steps, implementation details, as
well as unavailability of older versions of public search engines, the
results were encouraging. That is, with minor changes to the parameter
settings, we could reproduce results obtained by MerckKGaA, the team
ranked second in TREC 2016 CDS. We also confirmed the positive effect
of negation detection as implemented by some of the participating teams.

By using our platform, teams will be able to both report their methods
in a consistent way, and evaluate their results against a common
baseline. In the future, we aim to use this platform to systematically
evaluate a more diverse combination of search methods.

\paragraph{Acknowledgement}Authors would like to thank Falk Scholer (RMIT University) for his helpful feedback on the drafts of this paper. We are also grateful for the financial support from Decision Sciences Program at CSIRO, as well as constructive comments and feedback from our colleagues in Language and Social Computing Team, especially Stephen Wan.


\end{document}